\documentclass{emulateapj}

\catcode`\"=\active\let"=\"

\def\3{{\ss} }

\def\c12{{1\over 2}}

\def\plusplus{\raise 0.3ex\hbox{${\scriptstyle ++}$}{}}

\newcommand{\oversim}[2]{\protect{\mbox{\lower0.5ex\vbox{%
   \baselineskip=0pt\lineskip=0.2ex
   \ialign{$\mathsurround=0pt #1\hfil##\hfil$\crcr#2\crcr\sim\crcr}}}}} 
\begin{document}

\title{Unraveling the origin of the Monoceros Stellar Ring}
\author{Jorge Pe\~{n}arrubia\altaffilmark{1}, David Mart\'inez-Delgado\altaffilmark{2} \& Hans-Walter Rix\altaffilmark{3}}
\altaffiltext{1} {University of Victoria, Victoria, BC, Canada}
\altaffiltext{2} {Instituto de Astrof\'isica de Canarias, Tenerife, Spain}
\altaffiltext{3} {Max-Planck Institute fuer Astronomie, Heidelberg, Germany}

\begin{abstract}
We compare the predictions of the Pe\~narrubia et al. (2005) model for the Monoceros stellar ring around the Milky Way with new observational constraints that provide deeper insights on its origin. Recently, Grillmair (2006) found a coherent
sub-structure in a panoramic analysis of SDSS data, spanning  $5^\circ\times 65^\circ$ above 
the Galactic plane towards the anticenter. We show here that this structure strikingly matches the prior 
model predictions in projected position on the sky 
and the distance ($\sim 10$kpc) to within 20\%.  
Newly measured velocities within this sub-structure also match the model predictions perfectly.
This match suggests that the model assumptions are correct, namely, that the Monoceros Ring corresponds to a tidal stream wrapping the Milky Way that results from a single disruption event of a satellite ($M\sim 5\times 10^8 M_\odot$) with a prograde, low-inclination, low-eccentric orbit. 
Further support for the external origin of the Mon Ring comes from new metalicity measurements of the Mon Ring, which show that this system is much more metal poor than star clusters in the outer Galactic disk at the same radius, and independent detections of RRLyr over-densities in the area that are in excellent agreement with the model predictions.
That these observational data can be comprehensively explained by alternative disk heating scenarios seems not likely, but would need to be checked by detailed modeling.
\end{abstract}

\section{Introduction}\label{sec:int}

\begin{figure*}[!]
\vspace{10cm}
\includegraphics{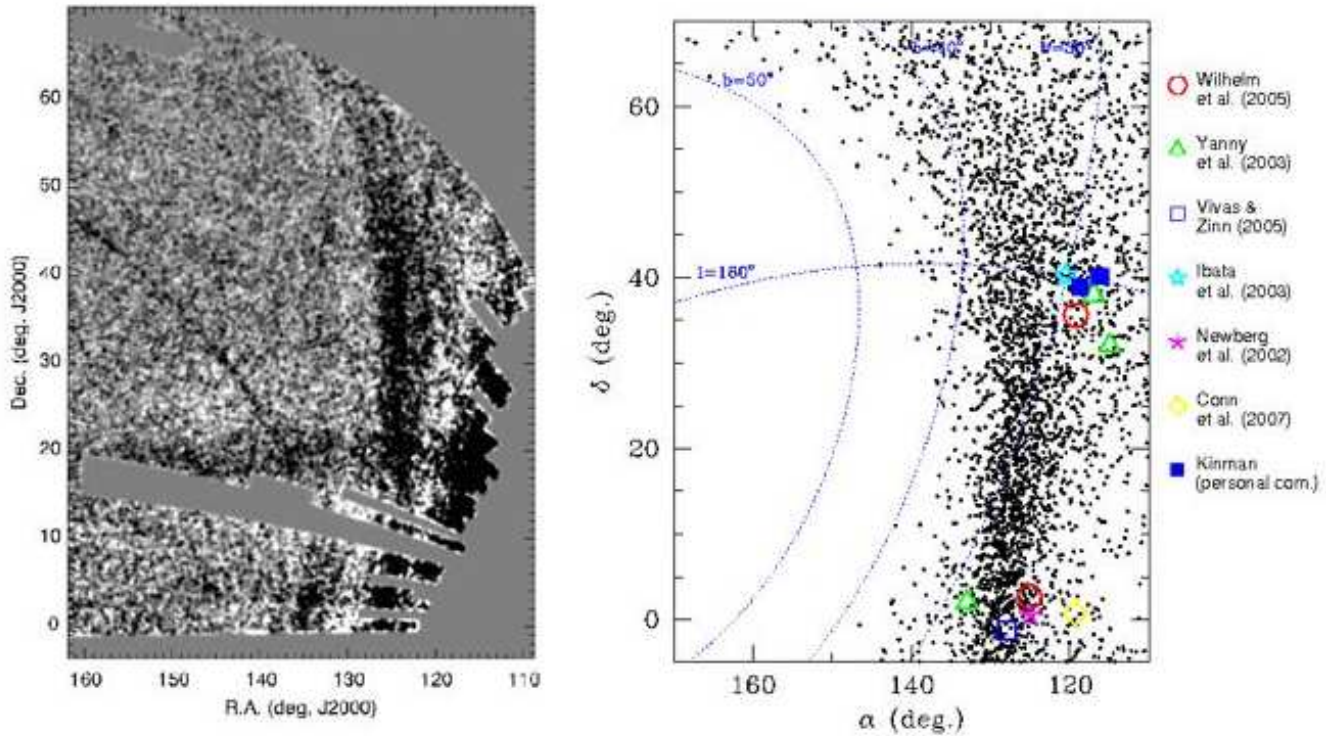}
\caption{{\em Left panel:} Stellar over-densities in the stellar density map of the Galactic anticenter region obtained by Grillmair (2006) using SDSS-I data. Darker areas denote higher surface brightness. The possible Mon stream detection spans from (R.A., del.)=(+126.4$^\circ$, -0.7$^\circ$) to (+133.9$^\circ$, +64.2$^\circ$). {\em Right panel:} Theoretical model of the {\em Monoceros tidal stream} (black particles, see Pe\~narrubia et al. 2005) in the Northern Galactic Cap. Note the excellent agreement between the Mon stream model and the coherent stellar structure visible in Fig.~1 at $\alpha\simeq +130^\circ$.}
\label{fig:model_proj}
\end{figure*} 

Hierarchical models exploring the formation of spiral galaxies (e.g. Abadi et al. 2003a,b; Governato et al. 2007 and references therein) show that galactic disks grow from inside out. In those models a large fraction of thick disk stars were actually brought into the disk by the merger of satellite galaxies and it is in the disk outer regions were the presence at a large number of coherent structures resulting from accretion events is expected.

Using SDSS data, Newberg et al. (2002) and Yanny et al. (2003) reported the detection of stellar over-densities at low galactic latitude that span about 100 degrees in the sky and locate approximately in the Anticenter direction. The whole structure was called the Monoceros (Mon) Stellar Ring or also (putatively) the Monoceros tidal stream. Follow-up  observations (Ibata et al. 2003) found that this structure of low metalicity stars surrounds the Galactic disk at Galactocentric distances from $\sim$ 8 kpc to $\sim 20$ kpc. Tracing this structure with 2MASS M giant stars,  Rocha-Pinto et al. (2003) suggested that this structure might be the fossil remnants of a merging dwarf galaxy.  Interestingly, Frinchaboy et al. (2004) and Crane et al. (2003) also suggested some nearby Galactic open and globular clusters with coordinated heliocentric radial velocities, indicating a possible common origin. 

Pe\~narrubia et al. (2005), hereinafter P05, collected all the positive detections at the time to build a comprehensive N-body Mon ring model under the assumption that its complex structure corresponds to several wraps of material that was stripped from a single satellite galaxy. After exploring more than $10^5$ orbits, they tightly constrained the orbital parameters of the stream progenitor: it has a low eccentricity, $e=0.10\pm0.05$ and orbital inclination with respect to disk plane, $i = 25^\circ \pm 5^\circ$, and the orbital sense is prograde relative to the Milky Way rotation. A similar result was also obtained by Martin et al. (2005). These kind of minor mergers are believed to be the building-blocks of 
the external disks of spiral galaxies and the recently discovered extended disk of M31 might have an origin similar to that of the Mon stream (Pe\~narrubia, McConnachie \& Babul 2006).

Since the progenitor was unknown, the model was expected to constrain its present position. Unfortunately, the observational data proved to be insufficient. 

Martin et al. (2004) detected an over-density of M-giant stars in Canis Major (CMa) located at low latitude $b=-8^\circ$, which they proposed to be the progenitor of the Mon stream. Several subsequent papers have argued against (Momany et al. 2004, 2006, Rocha-Pinto et al. 2006, Moitinho et al. 2006) and in favor (Mart\'inez-Delgado et al. 2005, Bellazzini et al. 2006) of this hypothesis without a general agreement. Recent follow-up, deeper observations carried out by Butler et al. (2006) and de Jong et al. (2007) in the CMa over-density region have not been conclusive about the nature of the CMa over-density but make it unlikely that it is the still-bound remnant of a satellite galaxy.

Crucially for the understanding the nature of the CMa over-density, Dinescu et al. (2005) measured the proper motions of the CMa young stellar population. Remarkably, the resulting CMa orbit matches, within one-sigma error, the orbit previously predicted by P05 for the Mon stream progenitor, which clearly suggests an association between both systems (but also see Moitinho et al. 2006 for a different interpretation).

Recently, independent analysis of large scale density maps from the SDSS data in the North Galactic Cap have shown the presence of a giant stellar structure in the direction of the Galactic Anticenter. Belokurov et al. (2006) detected stellar structures at low Galactic latitude that appeared to match the predictions of P05 Mon stream model in this area. Grillmair (2006), hereinafter G06, found that these stellar features form a coherent, narrow stellar stream that spans $5^\circ\times 65^\circ$ in parallel to the Milky Way disk (see Fig.1) and advocated for the detection of a new tidal stream of unknown origin.

The goal of this {\em Letter} is to contrast the properties of this newly discovered over-density with the predictions of P05 model and to explore the implications for the origin of the Mon stellar ring.


\section{Comparison with theoretical models}
\subsection{Over-densities in the Galactic Anticenter}
The left panel of Fig.~1 shows the image composition made by G06 using SDSS data in the direction of the Galactic Anticenter. Several over-densities can be readily identified. For example, it has been suggested (Belokurov et al. 2006, Fellhauer et al. 2006) that those visible in the whole Right Ascension interval at constant $\delta\sim+20^\circ$ and $+5^\circ$  correspond to the Sagittarius stream crossing the Galactic plane, although the lack of radial velocities prevent a certain association. A third, smaller over-density can be seen spanning from $(\alpha,\delta)=(+160^\circ,+45^\circ$) to $(+138^\circ,+20^\circ)$ which, owing to its small thickness, is likely associated to a Globular Cluster (G06).

\subsection{The Monoceros stream}
The Monoceros stream model constructed by P05 (black dots in Fig.~1) predicts the presence of a coherent tidal tail that spans between $-10^\circ\leq \delta\leq +70^\circ$  nearly in a parallel direction to the Milky Way disk (R.A. $\alpha\simeq +130^\circ$). In the past years, several stream detections in this area (thick dots) have been published. Unfortunately, their field coverage relative to the stream extension was small and it was unclear whether these detections were part of a single, extended system or they corresponded to different, independent stellar over-densities.

The recent image obtained by G06 show thus  a
 unprecedentedly large scale view of the coherent structure of the Monoceros stream. A comparison with Fig.~1 shows an excellent agreement with the model predictions, which successfully match the location and the width of the tidal stream. This last provides a further constraint on the mass of the stream progenitor (Johnston et al. 2001) which, owing to its slow orbital decay, was febly contraint by P05. The correct description of the stream width appears to validate P05 estimate $M\sim 5\times 10^8 M_\odot$.

\begin{figure}
\vspace{8.cm}
\includegraphics{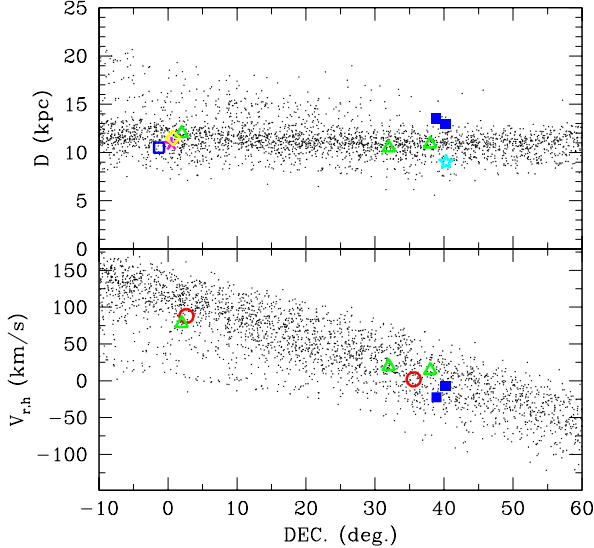}
\caption{Heliocentric distances (upper panel) and heliocentric radial velocities (bottom panel) predicted by the model of Pe\~narrubia et al. (2005) as a function of Declination. Individual detections are shown with thick dots following the notation of Fig.~1. According to G06, the detected stream locates at a distance of $D\simeq 8.9$ kpc, in good agreement with the model predictions.}
\label{fig:model_proj}
\end{figure} 

In addition to the projected position, G06 also measured heliocentric distances ($D$) along the stream, finding that $D\simeq 8.9$ kpc in the whole declination interval. This implies that its orientation is almost perpendicular to our line of sight. In the upper panel of Fig.~3 we show the heliocentric distances predicted by P05 model. We plot them as a function Declination, as all debris in this area locate at a similar Right Ascension ($\alpha\sim 130^\circ$). In a good agreement with previous distance measurements (thick dots) and the observational estimates of G06, the theoretical model predicts a single stream distance of $D\simeq 9$--11 kpc perpendicularly to the line of sight.

There are also radial velocity measurements of some of the SDSS stars 
that populate this over-density, which we denote with thick dots in the bottom panel. These appear in agreement with the gentle gradient of $V_{r,h}$ as a function of Declination predicted by the theoretical model, which goes from $V_{r,h}\sim 110$ km/s at $\delta=0^\circ$ to  $V_{r,h}\sim -70$ km/s at $\delta=+70^\circ$ and reflects the low orbital eccentricity of the Mon stream progenitor. 

New radial velocity measurements along the SDSS over-density are feasible with the present instrumentation and would put strong observational constraints on future model refinements.

\section{The stellar population of the Mon stream}
Further constraints on the nature of the Monoceros stream can be gleaned from the analysis of the stellar population in different stream pieces.

Several metalicity estimates in the Northern Galactic Cap area indicate that the composition of the Mon stream stars is largely dominated by a metal-poor population with no visible metalicity gradient. Wilhelm et al. (2005) use SDSS data to measure the spectroscopic abundance of main-sequence stream stars at $\delta\simeq +3^\circ$ and $\delta\simeq +36^\circ$ (see Fig.~1).
 In the southern plate, these authors estimate [Fe/H]$\sim -1.36\pm 0.45$, in agreement with the value of Yanny et al. (2003), [Fe/H]$\sim -1.6\pm 0.3$. Interestingly, in the northern field both groups obtain a metalicity value ([Fe/H]$\sim -1.39$) that is 
basically identical to the southern one -- despite being located $33^\circ$ away -- and confirms the homogeneous metalicity distribution in this piece of the Mon stream. 

 Photometric results match the spectroscopic ones. Analysing the CDM of these debris, G06 found a MS turnoff color in the range of $(g-r)\sim$ 0.23 to 0.25 (with an uncertainty of 0.02 mag), in  good agreement with those obtained by Yanny et al. (2003), $(g-r)_{o}\sim 0.26$. The similar main-sequence color distribution strengthen our conclusion than they are
part of the same stellar system. 


At the same time, in this region of the Galaxy the chemical composition of the satellite debris is clearly different to that of disk stars. This is illustrated in Fig.~3, which shows the metalicity measurements of old open clusters and Cepheid stars obtained by Carney et al. (2005) in a wide range of Galactocentric distances ($R_{\rm GC}$).  The metallicity of the Monoceros tidal debris stars, [Fe/H]$\simeq$-1.4, is significantly lower than the expectation for field stars in the outskirts region of the Galaxy, which has an averaged value of [Fe/H]$~\sim -0.48\pm 0.07$ dex. As Carney et al. (2005) state, the old disk stars enclosed in open clusters and the considerably younger Cepheids show a similar metalicity gradient. This, together with the fact that Cepheids have on average higher metalicities, suggests that the disk has grown radially over time. It also shows that the stars of Monoceros stream is unlikely associated with any stellar population of the Milky Way disk independent of its age. 


Using RR Lyrae stars as tracers of merger remnants, Vivas \& Zinn (2006) reported the discovery of an over-density in the QUEST (QUasar Equatorial Survey Team) survey in a range of ~7.5 deg that has a projected position (Fig.~1) and heliocentric distance (Fig.~2) consistent with those of the Mon stream. Unfortunately, radial velocity measurements (the model prediction is $100\leq V_{r,h}\leq 150$ km/s) are needed to stablish an association. Further evidence of the presence of an old stellar population in the stream are given by Kinman et al. ({\it per.comm.}), who detects RRLyr stars at $\simeq 40^\circ$ away from those reported by Vivas \& Zinn (2006), with a distance and radial velocity in excellent agreement with the P05 model predictions.
These independent detections confirm that the Monoceros stream contains stars as old as those of the Milky Way globular clusters ($\sim$ 12 Gyr). 
These ancient stars are not expected in the stellar population of the outer disk that formed {\it in situ}. According to cosmological simulations (e.g. Abadi et al. 2003a,b), most of the RRLyr stars present in the outer disk were accreted onto the Milky Way by the mergers of satellite galaxies.

On the other hand, this stream piece is considerably more metal-poor than its possible progenitor, the CMa over-density. CMa shows two different stellar components, a metal-rich one, $[Fe/H]\simeq -0.3$, that is possibly embedded in a more extended, metal-poor component with $[Fe/H]\simeq -1.0$ (de Jong et al. 2007), which has been observed in several dSphs in the Milky Way and M31 (see McConnachie et al. 2006 and references therein). 
In galaxies with differently segregated stellar populations, tidal forces are expected to affect more strongly the extended, metal-poor component (Pe\~narrubia, McConnachie \& Navarro 2007). Thus, if stars are stripped, a metalicity gradient along the tidal stream would naturally arise, with the remnant progenitor becoming systematically more metal-rich than the stripped stars. The metalicity difference between CMa and the Mon stream can be used as constraints on the age of different stream pieces. Using this information, P05 find that this part of the Mon stream became unbound to its progenitor around 2 Gyrs ago.

\begin{figure}
\vspace{8.cm}
\includegraphics{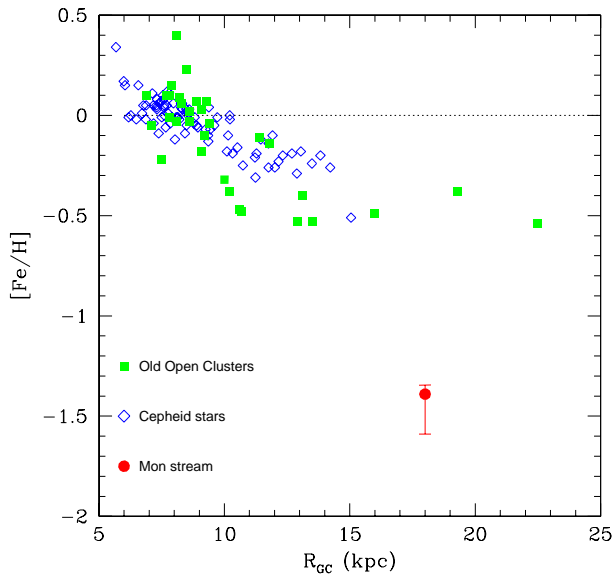}
\caption{Metalicity of old open clusters and Cepheids in the Milky Way vs their Galactocentric radius (from Carney et al. 2005) compared with the value obtained for the Mon stream in the Anticenter direction. Error bars denote the range of values obtained from independent measurements (see text). The errors of Carney et al. (2005) values are smaller than the dot size. Note that the chemical composition of the Mon stream in this area is clearly different to that of Milky Way stars.}
\label{fig:model_proj}
\end{figure} 

\section{Conclusions}
In this {\em Letter} we have compared the predictions of the P05 model with new
observational constraints. In particular, we considered the nature of the recently discovered stellar over-density arching 
as a coherent stream above the the Galactic Anticenter direction (G06). 
This system spans $5^\circ\times 65^\circ$ in the sky and at a nearly constant distance
of $\simeq 9$ kpc from the Sun. Comparing this with the predictions of the P05 Mnoceros Ring model
shows an excellent agreement, even though Grillmair had attributed this feature to a new ``stream''.
The P05 model startlingly matches the projected position, geometry, heliocentric distance and the available radial velocities 
at different positions along this over-density, strongly suggesting that this stream-like system is simply part of the Monoceros 
Ring. For the first time we have a wide, coherent 
picture of this tidal stream, previously known as a set of localized stellar over-densities dispersed over a large area of the sky.

The various low-latitude over-densities, termed the Monoceros Ring, has motivated alternative formation scenarios where the 
material constitutes a perturbed part of the stellar disk, for example,  a warp-like perturbation 
of the Milky Way disk (Momany et al. 2005),  an out-of plane spiral arm (Moitinho et al. 2006) 
or the remnants of an old disk that has been heated by accretion of multiple dark matter sub-halos (Kazantzidis et al,
{\it per. comm.}). 

The geometric analysis of the new observational constraints are clearly consistent
with an origin of the Monoceros Ring as a wrapped tidal tail of a disrupted satellite (P05).
The arguments about the stellar populations along the Monoceros Ring compiled here seem to favor an  
external origin of this system. 
The spectroscopic and photometric abundances clearly show a metalicity that is not compatible with
 that of the Cepheids and open clusters in this portion of the outer disk,
 but instead is similar to that exhibited by globular clusters in the Galactic halo. I.e it is dominated by a metal-poor population and significantly more metal poor than the thin and thick disk stars in this region of the Galaxy ($R_{GC}\simeq 18$--$20$ kpc). Furthermore, there are detections of RRLyr over-densities in this area (Vivas \& Zinn 2005, Kinman et al. 2007), with distances and radial velocities that are in excellent agreement with the P05 model predictions.
In a CDM cosmogony, galactic disks form small and grow from inside out. In this scenario, these ancient stars, as old as globular clusters, orbiting in outskirts of the disk cannot have formed {\it in situ} and the most reasonable explanation is that they are remnants of the accretion of external systems moving on low-eccentricity, low-inclination orbits (Abadi 2003a,b), similar therefore to that of the Mon Ring progenitor (P05).

This newly identified part of the Mon Ring is considerably more metal-poor than the CMa over-density,
 which is the best progenitor candidate at present (Martin et al. 2004). 
Metalicity variations along streams are expected from the results of Pe\~narrubia et al. (2007), who show that in galaxies that posses multiple stellar populations with different spatial segregations, tidal forces will affect more strongly the extended, metal-poor component. In the case of tidal stripping, a metalicity gradient would naturally arise and the remnant core would be on average systematically more metal-rich than the stripped stars.

The successful predictions made by P05 clearly strength the validity of their model assumptions, namely, that the Monoceros Ring
results from the disruption of an accreted dSph-like system ($M\sim 5\times 10^8 M_\odot$) 
with a prograde, low-inclination, low-eccentric orbit -- thus, similar to the orbit of thick disk stars -- 
and that the detections of the stream at different directions of the sky correspond to different orbital wraps.
 Whether
alternative models also can match these new constraints naturally, remains to be seen.


\vskip0.5cm
JP thanks Julio F. Navarro for financial support and useful comments. We are grateful to B. Carney and T. Kinman for kindly providing his data. JP also wants to thank K. Venn for very useful inputs on the metal abundances in the MW and dSphs. JP is grateful to A. McConnachie for constructive discussions.

{}

\end{document}